# Room-temperature vibrational properties of multiferroic MnWO$_4$ under quasi-hydrostatic compression up to 39 GPa

J. Ruiz-Fuertes[1,2,†], D. Errandonea[2], O. Gomis[3], A. Friedrich[1], and F. J. Manjón[4]

[1]Abteilung Kristallographie, Geowissenschaften, Goethe-Universität, 60438 Frankfurt am Main, Germany

[2]Departamento de Física Aplicada, Universitat de València, 46100 Burjassot, Valencia, Spain

[3]Centro de Tecnologías Físicas: Acústica, Materiales y Astrofísica, Universitat Politècnica de València, 46022 Valencia, Spain

[4]Instituto de Diseño para la Fabricación y Producción Automatizada, Universitat Politècnica de València, 46022 València, Spain

## Abstract

The multiferroic manganese tungstate (MnWO$_4$) has been studied by high-pressure Raman spectroscopy at room temperature under quasi-hydrostatic conditions up to 39.3 GPa. The low-pressure wolframite phase undergoes a phase transition at 25.7 GPa, a pressure around 8 GPa higher than that found in previous works which used less hydrostatic pressure-transmitting media. The pressure dependence of the Raman active modes of both the low- and high-pressure phases are reported and discussed comparing with the results available in the literature for MnWO$_4$ and related wolframites. A gradual pressure-induced phase transition from the low- to the high- pressure phase is suggested on the basis of the linear intensity decrease of the Raman mode with the lowest frequency up to the end of the phase transition.

† E-Mail address: ruiz-fuertes@kristall.uni-frankfurt.de

# 1. Introduction

Scintillators for dark matter search [1], semiconducting photoelectrodes for photoelectrolysis [2] or humidity sensors for meteorology [3] are some of the direct applications that wolframite-type compounds with chemical formula $A$WO$_4$ present. On top of that, an especially interesting case is the one with the divalent magnetic ion $A$ = Mn$^{2+}$, which shows three different antiferromagnetic phases below 13.7 K, with the intermediate one presenting an incommensurate magnetic structure able to lift the center of inversion in MnWO$_4$ originating a polar moment [4]. The high-pressure (HP) behavior of these materials has attracted a lot of attention in recent years [5,6], with the search of new structures with enhanced scintillating properties being the major cause. At ambient pressure, wolframites crystallize in a monoclinic structure with $Z$ = 2 and space group $P2/c$ [7]. A theoretical work has shown that the monoclinic ($P2/c$) wolframite structure is energetically competitive with a triclinic ($P\bar{1}$) one of CuWO$_4$-type [8]. On the other hand, experiments have demonstrated that non-uniform stress due to non-hydrostatic components can favor a structural transformation from the monoclinic to the triclinic structure [8,9]. This particular transformation induced by non-hydrostatic conditions has been observed experimentally in wolframites ZnWO$_4$ and MgWO$_4$ at around 17 GPa [8] and more recently at a similar pressure in MnWO$_4$ by Dai *et al.* [10]. In quasi-hydrostatic conditions, the absence of deviatory stresses has proved to allow these materials to retain the monoclinic wolframite structure up to pressures around 26 GPa [5,9] with a predicted doubled unit-cell and a symmetry increase for the HP phase that remains unsolved experimentally. Despite the technological and fundamental science interest that MnWO$_4$ generates, the study of its structural and optical properties under pressure in quasi-hydrostatic conditions has been limited to pressures below 10 GPa [6,11,12]. In this work we report a Raman spectroscopy study under quasi-hydrostatic conditions of the vibrational properties of MnWO$_4$ up to 39.3 GPa that shows an increase of 8 GPa in the structural stability of the wolframite phase with respect to previous works.

## 2. Experimental details

Our HP Raman experiments were performed at room temperature using diamond-anvil cells (DACs). To minimise the presence of deviatoric stresses that might compromise the phase diagram of MnWO$_4$, we have employed neon (Ne) as pressure-transmitting medium (PTM). Neon is known to guarantee quasi-hydrostatic conditions up to and above 30 GPa [13]. Powder samples were obtained from 1 mm$^3$ single crystals grown by the high-temperature solution method [14]. The measurements were conducted in backscattering geometry with a 632.81 nm HeNe laser line. A LabRam HR UV microRaman spectrometer with a 1200 grooves/mm grating and a 100 µm slit was employed together with a thermoelectric-cooled multichannel CCD detector ensuring a resolution better than 2 cm$^{-1}$. For the experiments, a 300-µm culet Boehler-Almax DAC and a tungsten gasket indented to 35 µm in thickness, and laser drilled with a 100 µm hole, were used. In all experiments the ruby fluorescence scale [15] was employed to determine pressure.

## 3. Results and discussion

**Figure 1** shows selected Raman spectra of MnWO$_4$ at different pressures up to 39.3 GPa. The point group of the low-pressure (LP) wolframite structure (C$_{2h}$) gives rise to 8 $A_g$ all-breathing modes and 10 $B_g$ modes that are Raman active. In our experiment we were able to follow the 18 Raman-active modes of the LP phase under pressure. We can observe a frequency shift to higher frequencies, some changes in the relative intensity of some of the modes or the loss of some of them, a peak width increase due to the inter-grain strain, and the frequency crossing of two other modes around 400 cm$^{-1}$ at 15 GPa already seen in other wolframites [5,9]. However, no evidence of a structural phase transition to a HP phase was found below 25.7 GPa. Above this pressure, 18 new Raman-active modes appear and coexist with those of the LP phase up to 35 GPa. The appearance of these new peaks at 25.7 GPa is interpreted as the onset of a structural phase transition that is only completed at around 35 GPa.

Low-frequency vibrational modes in wolframites are particularly interesting because they are due to $WO_6$-$MnO_6$ inter-octahedra motions and therefore are more sensitive to structural symmetry changes than the intra-octahedra stretching modes which are more related to local changes inside the octahedra [5]. In this respect, it is interesting to note the continuous intensity decrease with pressure of the lowest-frequency $B_g$ mode located at 94 cm$^{-1}$ at 3.6 GPa (see **Fig. 1**) which is no longer detected above 28.1 GPa when the phase transition is about to end. The intensity decrease of the lowest frequency $B_g$ mode in the LP phase of $MnWO_4$ with pressure has also been observed in a recent work by Dai et al. [10] in $MnWO_4$ but this mode does not show this behaviour in the other wolframites studied so far. It is well known that in the surroundings of a phase transition the intensity of the Raman active modes tends to drop as a consequence of the partial transformation of the crystal. However, a continuous intensity decrease of a particular mode is usually related to changes in the polarizability, $\alpha_p$. The intensity of a Raman mode is proportional to the square of the polarizability change with the normal mode coordinate $Q$: $I \propto (d\alpha_p/dQ)^2$. In **Fig. 2** we represent the pressure evolution of the square root of the relative intensity of the lowest-frequency Raman $B_g$ mode of $MnWO_4$ with respect to that of the highest-frequency $A_g$ mode $I(B_g)/I(A_g)$. Note that the highest frequency mode $A_g$ is the most intense one and its intensity almost does not change with pressure. Therefore, by normalizing the intensity of the lowest energy $B_g$ mode with respect the highest frequency $A_g$ mode we avoid uncontrolled effects that might affect the intensity of the modes. We have found that the square root of the relative intensity and therefore $d\alpha/dQ$, has a linear dependence with pressure. The intensity of the lowest energy $B_g$ mode is not accurately measurable in our experiment above 25.7 GPa, however if we extrapolate the linear tendency to higher pressure we observe that the mode would completely vanish at 35 GPa with the end of the phase transition. Although no much information can be obtained about the structural transformation from the variation of $d\alpha/dQ$ with pressure, any variation of the polarizability change of a mode must necessarily be related to the local symmetry of the block involved.

Considering the number of Raman-active modes and their relative intensities, the HP Raman spectra of $MnWO_4$ resemble very much those of the HP phases of other wolframites. This suggests that most wolframites undergo a phase transformation to the same HP phase. On the basis of this assumption, the linear intensity decrease of the lowest frequency $B_g$ mode in $MnWO_4$, though associated to the phase transition, cannot be the cause of the phase transformation since its pressure dependence is different for other wolframites. However, it clearly indicates that the mechanism of the phase transformation in $MnWO_4$ must be gradual due to the linear decrease of the intensity and its disappearance at the end of the phase transition.

Another interesting piece of information about the HP phase structure can be extracted from the highest-frequency Raman mode in wolframites, and in particular in $MnWO_4$. This all-breathing $A_g$ mode located at 895 cm$^{-1}$ at 3.6 GPa, is an internal $WO_6$ stretching mode and therefore its frequency is entirely dependent of the W-O bonding distances. Therefore, a frequency drop in this mode might be tentatively understood either as a W-O bond enlargement or a coordination increase, or both. However, direct structural information cannot be extracted from a spectroscopic technique. Therefore, the present results are calling for the performance of x-ray diffraction studies under high pressure in quasi-hydrostatic conditions. Another point to remark from **Fig. 1** is that the phase transition is completely reversible as can be shown with the Raman spectrum measured at 0.5 GPa after releasing the pressure.

In **Fig. 3** we present the pressure evolution of the 18 Raman modes of $MnWO_4$ for both the LP and the HP phases up to 39.3 GPa. Since most of the Raman modes evolve linearly with pressure, we have obtained the pressure coefficients ($d\omega/dP$) of them by means of linear fits. For those modes that have a quadratic behavior (219, 790, and 895 $cm^{-1}$ at 3.6 GPa) we have obtained the pressure coefficient from the LP linear part to allow comparison with previous works. HP and LP results are summarized in **Table I** together with the experimental results obtained in Refs. [10,16], and the calculated results presented in Ref. [9]. The agreement of the measured mode frequencies at ambient pressure for the LP phase in the three experiments is good; however, the behavior of some of the modes under pressure is rather dependent on the hydrostatic conditions. Hence, the pressure coefficients that we have obtained for some modes with neon as PTM differ considerably from those obtained by Dai et al. [10] without PTM or those obtained by Maczka et al. [16] in $Mn_{0.97}Fe_{0.03}WO_4$ using mineral oil Lujol as PTM. The agreement between the experimental and calculated [9] pressure coefficients is in general very good when comparing with the quasi-hydrostatic experiment, while there are important differences when theory is compared to non-hydrostatic results. It is worth to mention for example the lowest energy $B_g$ mode, which shifts its energy almost twice faster in non-hydrostatic media than in Ne whereas the $B_g$ mode at 512 $cm^{-1}$ at ambient pressure shifts twice slower. Both facts suggest that deviatoric stresses could strongly influence the high-pressure structural and vibrational behavior of wolframites as they do on related scheelite-type oxides [17, 18].

In **Table I** we also show the calculated grüneisen parameters $\gamma_i = (B_0/\omega)\cdot(d\omega/dP)$ for the LP phase of $MnWO_4$ according to our results. For this purpose we have employed the bulk modulus obtained in **Ref. 11** ($B_0 = 131(2)$ GPa). Since the multiferroic behavior of $MnWO_4$ was discovered, many works have been devoted to study its physical properties. However, most of those properties have been studied in the low-temperature range where the magnetic behavior shows up. One of those physical interesting properties is the thermal expansion that has been recently measured under pressure up to 1.4 GPa below 20 K [12]. However the value of the thermal expansion coefficient has not been reported so far in a higher temperature range and only a predicted value can be found in the bibliography at ambient conditions [19]. A way to obtain an estimation of its value near the room temperature range can be done if the grüneisen parameters of all the Raman active vibrational modes are known. The average grüneisen parameter $\gamma_{av}$ is related to the total molar heat capacity at constant volume, $C_v = \sum C_i$ and the linear thermal expansion coefficient $\alpha$ according to $\gamma_{av} = 3\alpha V_m B_0/C_v$ [20, 21], where $V_m$ is the molar volume and $C_i$ is the contribution to the total molar heat capacity of the mode $i$. The quantity $\gamma_{av}$ can be also obtained as $\gamma_{av} = \sum C_i\gamma_i/C_v$. The molar heat capacity of a single mode is calculated according to the formula:

$$C_i = R\frac{(E_i/k_BT)^2 e^{E_i/k_BT}}{(e^{E_i/k_BT}-1)^2}$$

where $E_i = h\nu_i$, and $R$ is the ideal gas constant. For $MnWO_4$ the molar unit-cell volume at ambient pressure is $V_m = 4.22\times10^{-5}$ m$^3$ mol$^{-1}$, and therefore we obtain a value for $C_v$ of 110 J mol$^{-1}$ K$^{-1}$ and for the average grüneisen parameter $\gamma_{av}$ of 0.69. Finally we obtain that the average thermal expansion coefficient $\alpha$ takes a value of $4.59\times10^{-6}$ K$^{-1}$. This value is comparable with the previously predicted average value of $5.7\times10^{-6}$ K$^{-1}$ since in our determination we are including only the contribution from the Raman active modes since no information about the grüneisen parameters of the infrared active modes currently exists. This gives rise to a necessary underestimation of the value of α in our calculation. Anyway the obtained value is also comparable to the calculated α, considering only the Raman modes contribution, for other tungsten oxides like the scheelite-structured $BaWO_4$ [22].

Concerning the HP phase, most of the Raman-active modes of the HP phase show smaller pressure coefficients than those of the LP phase, as observed in other wolframites. Since higher pressure phases are denser, they normally show an increased bulk modulus $B_0$ in comparison to their LP counterparts. This may explain the decrease in the pressure coefficients of the modes in the HP phase.

## 4. Concluding remarks

We have found with a Raman spectroscopy study under quasi-hydrostatic conditions that the structural stability of the wolframite phase in $MnWO_4$ is very sensitive to non-hydrostatic conditions and as happens with other wolframites, the stability range is extended to around 26 GPa when a quasi-hydrostatic pressure transmitting medium is used. The compound undergoes a phase transition to an unknown phase at 25.7 GPa that is fully completed at 35 GPa. The pressure dependence of the polarizability variation for the lowest frequency Raman mode $B_g$ has been obtained in the LP phase indicating that the phase transition must be gradual. Furthermore, the frequency drop of the highest frequency mode in the transformation to the HP phase may be interpreted as a change of the W-O distances in the polyhedron. An x-ray diffraction study in quasi-hydrostatic condition is needed and currently in progress to solve the HP phase and understand the transformation mechanism.

**Acknowledgements**

This work has been supported by the Spanish government under grant MAT2010-21270-C04-01/04, by MALTA Consolider Ingenio 2010 Project (CSD2007-00045), by Generalitat Valenciana (GVA-ACOMP-2013-1012), and by the Vicerrectorado de Investigación y Desarrollo of the Universidad Politécnica de Valencia (UPV2011-0914 PAID-05-11 and UPV2011-0966 PAID-06-11). We thank Prof. Gospodinov, Institute of Scintillating Materials in Ukraine, for providing us high-quality $MnWO_4$ single crystals. J.R.-F. thanks the Alexander von Humboldt Foundation for a postdoctoral fellowship. A.F. acknowledges support from the Germany Research foundation within the priority program SPP1236 (project FR-2491/2-1). The use of the SPP1236 central facility in Frankfurt is acknowledged.

**Bibliography**

[1] V. B. Mikhailik, H. Kraus, V. Kaputstyanyk, M. Panasyuk, Y. Prots, V. Tsybulskyi and L. Vasylechko, J. Phys.: Condens Matter **20**, 365219 (2008).

[2] M.A. Buttler, J. Appl. Phys. **48**, 1914 (1977).

[3] E. Traversa, Sens. Actuators B **23**, 135 (1995).

[4] K. Taniguchi, N. Abe, T. Takenobu, Y. Iwasa, and T. Arima, Phys. Rev. Lett. **97**, 097203 (2006).

[5] D. Errandonea, F. J. Manjón, N. Garro, P. Rodríguez-Hernández, S. Radescu, A. Mujica, A. Muñoz, and C. Y. Tu, Phys. Rev. B **78**, 054116 (2008).

**[6]** J. Ruiz-Fuertes, S. López-Moreno, J. López-Solano, D. Errandonea, A. Segura, R. Lacomba-Perales, A. Muñoz, S. Radescu, P. Rodríguez-Hernández, M. Gospodinov, L. L. Nagornaya, and C. Y. Tu, Phys. Rev. B **86**, 125202 (2012).

**[7]** A. W. Sleight, Acta Cryst. B **28**, 2899 (1972).

**[8]** J. Ruiz-Fuertes, S. López-Moreno, D. Errandonea, J. Pellicer-Porres, R. Lacomba-Perales, A. Segura, P. Rodríguez-Hernández, A. Muñoz, A. H. Romero and J. González, J. Appl. Phys. **107**, 083506 (2010).

**[9]** J. Ruiz-Fuertes, D. Errandonea, S. López-Moreno, J. González, O. Gomis, R. Vilaplana, F. J. Manjón, A. Muñoz, P. Rodríguez-Hernández, A. Friedrich, I. A. Tupitsyna, and L. L. Nagornaya, Phys. Rev. B **83**, 214112 (2011).

**[10]** R.C. Dai, X. Ding, Z.P. Wang, Z.M. Zhang, Chem. Phys. Lett. **586**, 76 (2013).

**[11]** J. Macavei and H. Schulz, Z. Kristallogr. **207**, 193 (1993).

**[12]** R. P. Chaudhury, F. Yen, C. R. de la Cruz, B. Lorenz, Y. Q. Wang, Y. Y. Sun, and C. W. Chu, Physica B **403**, 1428 (2008).

**[13]** S. Klotz, J.-C. Chervin, P. Munsch and G. Le Marchand, J. Phys. D: Appl. Phys. **42** 075413 (2009).

**[14]** M. N. Iliev, M. M. Gospodinov, and A. P. Litvinchuk, Phys. Rev. B **80**, 212302 (2009).

**[15]** H. K. Mao, J. Xu, and P. M. Bell, J. Geophys. Res. **91**, 4673 (1986).

**[16]** M. Mączka, M. Ptak, K. Pereira da Silva, P. T. C. Freire, and J. Hanuza, J. Phys. Condens. Matter **24**, 345403 (2012).

**[17]** D. Errandonea , L. Gracia , R. Lacomba-Perales , A. Polian  and J. C. Chervin, J. Appl. Phys. **113**, 123510 (2013).

**[18]** O. Gomis, J. A. Sans, R. Lacomba-Perales, D. Errandonea, Y. Meng, J. C. Chervin, and A. Polian, Phys. Rev. B **86**, 054121 (2012).

**[19]** H. Li, S. Zhou, and S. Zhang, J. Sol. State. Chem. **180**, 589 (2007).

**[20]** N. W. Ashkroft and N. D. Mermin, *Solid State Physics* (W. B. Saunders Company, Philadelphia, 1976), Chap. 25, p. 493.

**[21]** A. M. Hofmeister and H. K. Mao, Proc. Natl. Acad. Sci. **99**, 559 (2002).

**[22]** R. Lacomba-Perales, D. Martínez-García, D. Errandonea, Y. Le Godec, J. Philippe, and G. Morard, High Pressure Research **29**, 76 (2009).

**Table I.** Comparison of the measured Raman active modes obtained in this work at ambient pressure and those obtained by Dai et al. [10], and Maczka et al. [16] ($Mn_{0.97}Fe_{0.03}WO_4$), as well as the calculated modes [9] for the LP phase of $MnWO_4$ together with their pressure coefficients. The Raman active modes of the HP phase at 34 GPa with their pressure coefficients are also given.

| Symm. | Dai *et al.* [10] MnWO$_4$ | | Maczka *et al.* [16] Mn$_{0.97}$Fe$_{0.03}$WO$_4$ | | Present experiment MnWO$_4$ | | | Calculations [9] MnWO$_4$ | | Present experiment HP phase MnWO$_4$ | |
|---|---|---|---|---|---|---|---|---|---|---|---|
| | $\omega$ (cm$^{-1}$) | d$\omega$/dP (cm$^{-1}$GPa$^{-1}$) | $\omega$ (cm$^{-1}$) | d$\omega$/dP (cm$^{-1}$GPa$^{-1}$) | $\omega$ (cm$^{-1}$) | d$\omega$/dP (cm$^{-1}$GPa$^{-1}$) | $\gamma$ | $\omega$ (cm$^{-1}$) | d$\omega$/dP (cm$^{-1}$GPa$^{-1}$) | $\omega$ (cm$^{-1}$) | d$\omega$/dP (cm$^{-1}$GPa$^{-1}$) |
| *B$_g$* | 89 | 1.04 | | | 89 | 0.73 | 1.07 | 95 | 0.78 | 146 | 0.80 |
| *A$_g$* | 129 | 0.21 | | | 129 | 0.02 | 0.02 | 129 | -0.06 | 186 | 0.09 |
| *B$_g$* | | | 159 | 0.73 | 160 | 0.22 | 0.18 | 165 | 0.27 | 196 | 1.78 |
| *B$_g$* | 166 | 0.83 | 165 | 1.34 | 166 | 0.78 | 0.62 | 171 | 0.54 | 217 | 1.73 |
| *B$_g$* | | | | | 177 | 1.03 | 0.76 | 183 | 0.72 | 242 | 1.58 |
| *A$_g$* | 207 | 2.55 | | | 206 | 2.01 | 1.28 | 226 | 2.19 | 292 | 0.78 |
| *B$_g$* | | | | | 272 | 2.03 | 0.98 | 278 | 1.82 | 314 | 1.34 |
| *A$_g$* | 259 | 0.4 | 257 | 0.32 | 258 | 0.30 | 0.15 | 264 | 0.34 | 370 | 2.02 |
| *B$_g$* | | | 292 | 1.84 | 294 | 2.02 | 0.90 | 296 | 2.72 | 388 | 2.60 |
| *A$_g$* | 327 | 2.24 | 325 | 2.05 | 327 | 1.50 | 0.60 | 338 | 2.4 | 446 | 2.46 |
| *B$_g$* | | | 357 | 3.82 | 356 | 4.09 | 1.51 | 373 | 4.6 | 495 | 1.50 |
| *A$_g$* | 398 | 1.7 | 397 | 1.6 | 397 | 1.69 | 0.56 | 389 | 1.71 | 511 | 1.34 |
| *B$_g$* | 510 | 1.39 | 510 | 2.29 | 512 | 2.86 | 0.73 | 509 | 2.93 | 586 | 1.19 |
| *A$_g$* | 545 | 2.72 | 545 | 2.62 | 545 | 2.39 | 0.58 | 548 | 2.77 | 676 | 1.86 |
| *B$_g$* | 674 | 4.08 | 674 | 3.56 | 674 | 4.20 | 0.82 | 662 | 3.79 | 710 | 1.46 |
| *A$_g$* | 699 | 2.69 | 699 | 2.95 | 698 | 3.08 | 0.58 | 694 | 2.75 | 784 | 1.00 |
| *B$_g$* | 774 | 0.70 | 774 | 4.07 | 774 | 3.58 | 0.61 | 775 | 3.54 | 810 | 3.26 |
| *A$_g$* | 885 | 1.84 | 884 | 2.48 | 885 | 1.63 | 0.24 | 858 | 1.82 | 871 | 0.69 |

**Figure Captions:**

**Figure 1.** (Color online) Raman spectra of $MnWO_4$ at selected pressures. Black and red ticks locate the maxima of the Raman modes of the low-pressure phase at 3.6 and 25.7 GPa, respectively, whereas green and blue ticks do it for the high-pressure phase at 25.7 and 37.4 GPa, respectively. All spectra are measured upon pressure increase with the exception of the one at the top denoted by (r), which corresponds to pressure release. The symmetry of some of the modes is depicted.

**Figure 2.** (Color Online) Pressure dependence of the square root of the relative intensity of the lowest frequency $B_g$ mode with respect to the intensity of the highest frequency $A_g$ mode of the $MnWO_4$ low-pressure phase.

**Figure 3.** (Color Online) Pressure dependence of the Raman mode frequencies of the wolframite low-pressure (solid symbols) and high-pressure (empty symbols) phases of $MnWO_4$. Both linear (continuous) and quadratic (dotted) fittings are in red lines. The vertical line indicates an estimation of the onset of the phase transition.

**Figure 1.**

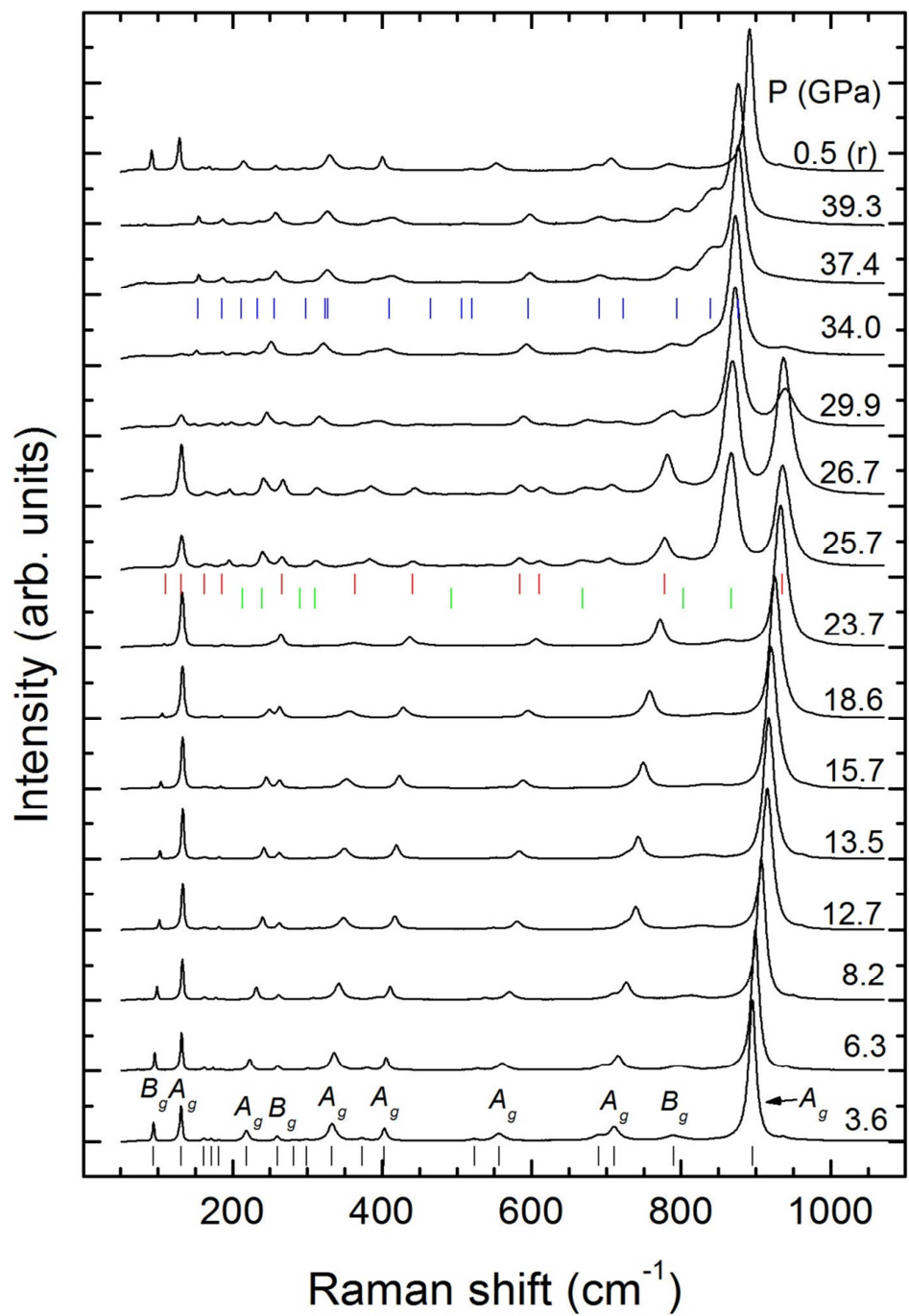

**Figure 2.**

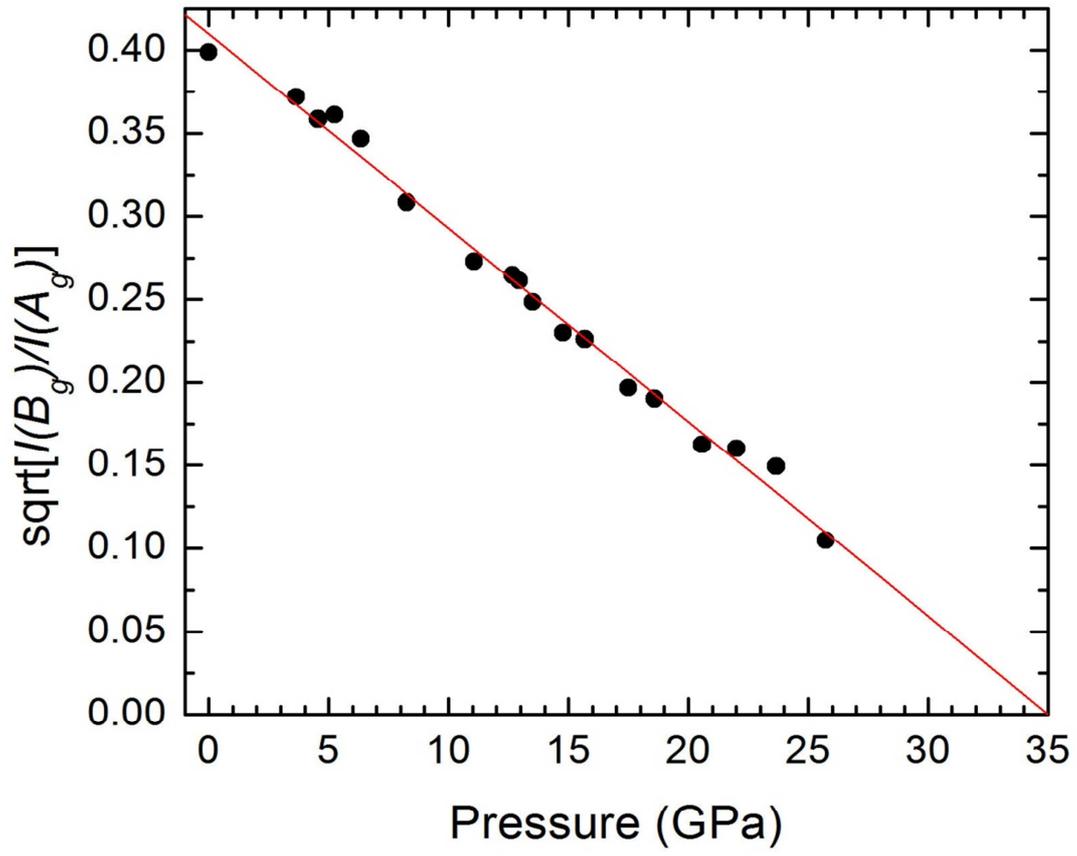

**Figure 3.**

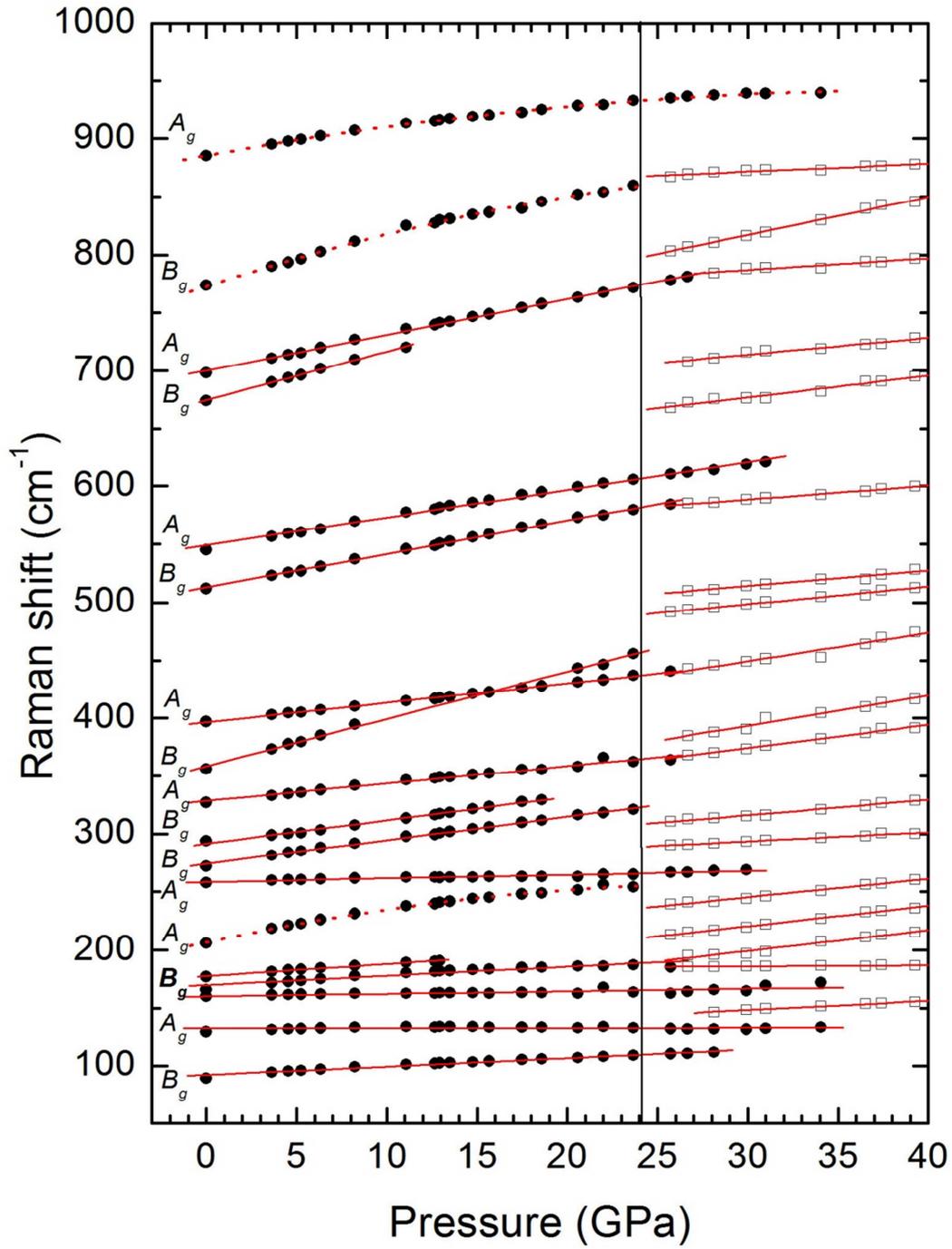